# A fundamental test for physics: the galactic supermassive obscure bodies

Maurizio Michelini
ENEA, Casaccia Research Centre, via Anguillarese 301, Rome, Italy

**Abstract**
The problem of the enormous gravitational effects observed by the astronomers in several galactic nuclei in connection with obscure bodies appears puzzling. These bodies, denouncing up to some million of gravitational Sun masses, are too large to be neutron stars generated by supernovae of huge primeval stars, but are too small to be considered a new kind of obscure quasi stellar objects. This contradiction represents a fundamental test which may destroy the established gravitational theories. We present a new physical paradigm which substitutes the old "gravitational mass" guessed by Newton more than tree centuries ago.

**Introduction**

The lucky observation with the X-ray Newton satellite (ESA) in a distant galaxy of three luminous bodies (27 hours period) orbiting around a "black hole" gave recently the possibility to calculate a gravitational mass around 300,000 Sun masses [1]. Previously, an intermediate "black hole" of 1300 Sun masses was found observing a rotating cluster of seven stars within the centre of our galaxy [2]. In this region it was known to reside a huge obscure galactic mass, firstly estimated around $2.6 \times 10^6$ solar masses [3] and later calculated $3.7 \times 10^6$ solar masses through a ten-years observation of an orbiting star [4]. D.Figer recalled [5] that in the Arches cluster the probability there are stars heavier than 150 solar masses equals $10^{-8}$. In the same Nature's issue, P.Kroupa [6] pointed out that the Arches cluster is populous enough to infer the existence of stars as massive as 150 sun masses. Since they are totally absent, he concluded that they are unable to form.

Here we discuss the origin of the huge obscure bodies which upset all observations of neutron stars accounting for the masses of their stellar progenitors.

According to the current ideas, the obscure bodies originate from the slow cooling and contraction of the neutron stars formed in the implosive collapse (supernovae) of massive stars. The masses of the obscure neutron stars in binary systems (where the swallowing of matter through the accretion disk from the luminous companion has been observed by the X-ray emission) are in the range 3-21 solar masses [7,8,9] which agree with the stellar progenitors.

The huge masses of several galactic obscure bodies is currently analysed from the standpoint of primeval supermassive stars which underwent a supernova or assuming that the mechanism of mass accretion ascertained in binary stars might be extended to the galactic system.

As to the first conjecture, R.Larson and V.Bromm [10] suggest that during the formation of galaxies there was the possibility that early stars could form up to 900-1000 Sun masses. A preceding work of Abel, Bryan and Norman found figures not exceeding 300 solar masses. So far, the results do not validate the first hypothesis.

Models of gas accretion by supermassive obscure objects have been realised observing many active galactic nuclei (AGN) which present an impulse of star formation, since the gas accretion often happens (but not always) together with the formation of millions of new stars. The two phenomena work in *competition*, because both are fuelled by the same distributed gas. These models predict that the obscure bodies may grow up to $10^8 \div 10^9$ Sun masses [11] with a mechanism as efficient as that observed in binary systems. It seems however that these high accretion rates have to be confined to the epoch *before* the formation of the stars, since their radiation blow reduced strongly the galactic gas density, thus producing a substantial limitation to the growth of the obscure bodies.

## 2- A simple model of gas accretion upon isolated, nearly obscure, neutron stars

To give an idea of the gas accretion rate upon a rather massive neutron star with mass $M_o$ which formed within the primeval galaxy, a simple model can be adopted considering that the gravitational mass-transfer of galactic gas through a spherical surface at a distance $r$

(1) $\qquad dM/dt \cong 4\pi r^2 u(r) \delta(r)$

appears to be a steady phenomenon when $r$ is sufficiently larger respect to the region of the X-ray emission by falling matter. In fact the net centripetal velocity of the molecules results $u(r) \approx \frac{1}{2}(GM/r^2)\tau$ where the time of flight $\tau = l_g/v_g$ is the ratio between the gas mean free path $l_g = m_g/\sigma_g \delta(r)$ and the molecule thermal velocity $v_g = (2kT/m)^{1/2}$. This model is valid as far as the velocity $u(r) < v_g$, that is at distances $r > 3 \times 10^{11}$ where the temperature of the ancient galactic gas is put by the astronomers around 100°K. Substituting these quantities in eq.(1) one gets an expression which, in the above assumptions, does not depends on $r$

(2) $\qquad dM/dt \approx 2\pi(m_g/\sigma_g) GM/v_g$.

This expression is not accurate to calculate the mass falling on the obscure star. However in the assumption of a steady mass-transfer up to the region of the X-ray emission, the mass accretion rate can be roughly calculated through eq.(2).

Integrating over time we obtain the mass increment during the time $\Delta t$

(3) $\qquad M/M_o \approx \exp[2\pi(m_g/\sigma_g) G\Delta t/v_g]$

Supposing the gas is hydrogen, then $(m_g/\sigma_g) \approx 10^{-7}$ and $v_g \approx 10^3$. Assuming also that $\Delta t$ equals the current age $t_U$ of the universe, the mass increment results $M/M_o \approx 1\%$.

However considering the presence of the cosmic powder (size $2 \times 10^{-8}$ according to Schalen) constituting the so called obscure nebulae, the ratio $(m_p/\sigma_p)$ takes figures around $4 \times 10^{-5}$. The powder velocity $v_p$ is governed by gravitation as well as the net centripetal velocity, thus complicating the calculation. The mass increment $M/M_o$ may probably reach figures up to few hundreds, depending on the amount of powder within the hydrogen.

An alternative mean of investigation is suggested by the fact that the galactic gas needs a long time $t_g$ to reach the neutron star from the distance $R_i$

(4) $\qquad t_g = \int_{ro}^{Ri} dr/u(r) \approx \int_{ro}^{Ri} [2 r^2 \delta(r) v_g / (m_g/\sigma_g) GM] dr \approx (M_g/M) (v_g/2\pi(m_g/\sigma_g) G)$

where $M_g$ is the mass of the displaced gas. Since this time must be lower than the assumed age $t_U$ of the expanding universe, the above equation puts a limit to the ratio

(5) $\qquad (M_g/M) \leq 2\pi t_U (m_g/\sigma_g) G/v_g$.

Substituting $(m_g/\sigma_g)$ and the velocity $v_g$ related to the hydrogen gas, we obtain a figure $(M_g/M) \leq 0.012$. Substituting also the figures related to the cosmic powder, the ratio $(M_g/M)$ may reach figures up to some units.

These results reinforce the conviction that either the age of the universe is greater than currently supposed, or the huge gravity observed in connection with several galactic obscure bodies is originate by phenomena totally unknown at present.

## 2 – Critique of the gravitational mass

Here we present a theory which accounts for the huge gravity of the observed bodies without producing troublesome effects on astronomy or physics. In fact the difficulty of explaining the observed huge masses may be due to a misconception linked to the gravitational theories based on the universal gravitational constant $G$.

We think that in the mentioned reliable astronomical measurements, the *gravity* acting on the nearest luminous bodies *really* attains 1.300 or 300.000 or 3.7 million times the standard $G$, but the mass of the obscure bodies (probably ancient neutron stars) does not necessarily take figures which disagree with the stellar progenitors.

May these contrasting features be both correct ?

A rational answer to this question requires to renounce the reassuring concept of gravitational mass, the paradigm proposed by Newton three centuries ago, but never legitimate by the epistemology investigation. The advent of the electromagnetic field theory (Maxwell) gave us a complete and rational description of the interaction between two charges by means of waves.

The gravitational interaction has been object of many experimental and theoretical attempts trying to set it within the same conceptual frame of the electromagnetic field. So far the existence of the mysterious "graviton", the field carrying particle, was not proved.

Besides, the origin of the very strong inertial forces remains mysterious. What physical mechanism originates, within times so small to escape any measurement, the strong forces which fragment a rotating steel wheel? The attempt to explain the inertial forces through the gravitational field of the distant masses of the universe (Mach's principle) seems to day only an obsolete conjecture. Conversely, the objections advanced in the past that G.R. does not entirely fulfil the *strong* equivalence principle (inertia and gravitation descend both from a unique physical phenomenon) are to day reappearing.

The crisis of the standard cosmology shows the last sign when the old cosmological constant is reintroduced [12] to reconcile the standard relativistic model and the cosmological observations.

## 3 - A new paradigm explaining the huge masses of the galactic obscure bodies

This physical paradigm claims that the gravitational force originates not from the *drawing* gravitational masses, but from the flux of cosmic quanta which *push* each other the masses.

The space is filled by a flux $\phi_o$ of very small quanta - the wavelength $\lambda_o$ equals the Planck's length - which easily penetrate the ordinary matter, travelling across the cosmic space.

As a consequence [13] the energy density of the void is $э_o = \phi_o E_o/c$, whose high figure was firstly estimated in 1967 by Y.Zeldovich.

*a) The origin of the mass*

First of all, the new paradigm establishes a theory of the *mass origin*, since any particle presents an interaction cross-section $\sigma$ which is proportional to the mass, so the ratio $(\sigma/m) = A_o$ is constant for any particle. In this frame, the unlimited gravitational collapse vanishes since the density of matter cannot exceed a certain limit [13] linked to the cross-section offered by the particles.

*b) The relativistic mechanics*

The new paradigm satisfies the laws of relativistic mechanics, explaining the increase of mass when the speed increases and the equivalence between mass and energy. In other words the space within which the masses move is no more a mathematical entity (although deformed by the presence of masses, as in G.R.), but a *physical* reality which imposes on the masses the laws of motion and energy conservation [13]. The flux $\phi_o$, coming isotropically from the space, collides upon an isolated nucleon with a high number of *simultaneous* collisions. The quanta gives no energy to the particle because any quantum hitting with the momentum $E_o/c$ is neutralised by an other quantum with opposite momentum.

Being $E_o = h_o \nu_o$ the quantum energy, a moving particle receives through the relativistic Doppler effect a forward momentum $q_f$ and a backward momentum $q_b$ which realise a dynamical equilibrium between the particle and the quanta. From these premises it has been demonstrated [13] that a *free* particle moving with velocity **v** respect to an inertial frame (i.e. respect to the large masses floating within the bubble of the universe filled by the cosmic flux) shows the relativistic momentum

(6) $\qquad\qquad\qquad\qquad q = q_f - q_b = m_o v / (1 - v^2/c^2)^{1/2}$

when the rest-mass $m_o = E_o \sigma \phi_o / v_o c^2$ is linked to the characteristics of the quanta.
This implies that the inertial forces arising from the time derivative of the relativistic momentum do not originate from the space, but from the interaction of the matter with the cosmic flux.

*c) The equivalence principle and the gravitation*

In accord with the "*strong*" principle of equivalence, the interaction of cosmic quanta with matter originates *both* gravitation and inertia.

Let's consider a star of mass $M$ (radius $R$) and a particle $m$ at distance $r$. Coming from the stellar mass, the beam of flux $\gamma(r)\phi_o$ comprised in the shielding angle $\gamma(r) = \pi R^2/4\pi r^2$ hits the particle with quanta of momentum $E_n/c$, where the reduced energy $E_n = E_o/(1+nK_o)$, derived from the Compton's equations, depend on the $n$ collisions which, on the average, the quanta underwent within the star.

By consequence the particle experiences a centripetal force due to the balance between the above beam and the opposite beam of quanta with momentum $E_o/c$ coming from the free space

$$(7) \quad f(r) = \sigma \phi_o (E_o - E_n) R^2 / 4 c r^2 .$$

We shall call "gravitational" this force although the two masses do not draw each other, but are *pushed* each toward the other. Substituting the expression of $E_n$ into eq.(7) one gets (being always $nK_o \ll 1$)

$$(8) \quad f(r) \cong \sigma n K_o \phi_o E_o R^2 / 4 c r^2 .$$

Being $l = m/\sigma\delta$ the mean free path within a mass with density $\delta$, the optical thickness of a star is $a = A_o M/\pi R^2$. For instance the Earth has an optical thickness $a \approx 2.2$, the Sun shows $a \approx 62$. The white dwarf stars show an optical thickness in the range $10^5 \div 10^6$. Substituting the optical thickness in eq.(8) we obtain

$$(9) \quad f(r) \cong (n/a) [K_o \phi_o E_o A_o^2 / 4\pi c] M m / r^2$$

which, putting the expression in brackets equal to the universal constant $G$, becomes the generalised gravitational equation

$$(10) \quad f(n,r) \cong (n/a) G M m / r^2$$

where the force depends not only on the geometry, but also on the energetic level of the interaction, expressed by the average number of collisions within the dense mass.

The ratio $(n/a)$ is called the *gravity factor* since it multiplies the newtonian constant. It results greater than 1 because while $a$ represent the collision number of a hypothetical quantum crossing the mass in a straight line, conversely $n$ is the number of *real* collisions (which produce always a certain deviation) suffered on the average by the crossing quanta. By consequence, if the mass is dense enough, the quantum may do a long zigzag before to exit.

As far as the optical thickness of the mass is not too high (for instance less than $10^5$), the relationship $n \cong a$ is satisfied. Hence the force defined by eq.(10) coincides with the newtonian force.

*d) The supergravity of the very dense stars*

This phenomenon originates when the cosmic quanta $E_n$ are sensibly weakened after crossing the dense mass. The number $n$ can be calculated imposing the stability of the star, i.e. the balance between the gravitational pressure and the gas pressure. In the case of ordinary stars made of atoms and molecules, the gas pressure is given by the equation of the ideal gases. By consequence the gravity factor results equal to unity (newtonian gravitation).

For the white dwarfs and neutron stars the gas pressure depends on the density of the degenerate matter $p_d = \pi^2 \hbar^2 \delta^{5/3} / 5 m_e m^{5/3}$. In this case one gets, at equilibrium conditions, an average number of collisions

$$n_{eq} \cong 2.26 \times 10^7 \delta (1+\chi) / M^{1/3}$$

where $\chi$ is the ratio between the radiation pressure and the gas pressure and $M$ is the star mass. Dividing by the optical thickness, we obtain the gravity factor depending on the average star density

$$(11) \quad (n_{eq}/a) \cong 5.8 \times 10^{17} \delta^{1/3} (1+\chi) / M^{2/3} .$$

For instance, substituting the average density of a white dwarfs, one gets a gravity factor equal to some units. Substituting the average density of ordinary neutron stars one obtains $(n_{eq}/a) \approx 10^3(1+\chi)$, whereas substituting the very high density of a star at final contraction [13] the gravity factor grows up to about $2 \times 10^6(1+\chi)$.

The incredible large masses of the galactic obscure bodies have to be interpreted as non newtonian gravitational forces related to the weakened cosmic quanta $E_n$ leaving the neutron star after doing many collisions. Any dense star shows a proper gravitational constant $G^* = (n_{eq}/a) G$ since the theory predicts a local (non-universal) gravity.

## 4 – Other conceptual innovations of the theory

An other problem which finds solution is the permanence of the gravitational field/force outside the obscure dense bodies, which conversely reduce the energy of the electromagnetic waves, causing the gravitational redshift. In the classical gravitation theory this fact puts a problem, since it was not clear why the obscure bodies brake the electromagnetic waves, but does not alter the gravitational waves.

It was argued that gravitation generates a different kind of field respect to the electromagnetism. Now we can see this difference. The cosmic quanta work on a scale extremely small respect to the electromagnetism: their wavelength is $10^{-30} \div 10^{-20}$ times smaller than the usual photon wavelength.

- *The missing mass in the spiral galaxies*

The observations show that the total mass of the stars $\Sigma M_i(r)$ along the galactic radius is not sufficient to bind the stars [14], since the observed rotation velocities are practically constant along the spiral arms whereas the Newton's velocities $v^2(r) = GM(r)/r$ reduce.

This discrepancy has been explained introducing a sort of "hidden" gravitational mass.

A concrete explanation may be found taking into account the gravity factor $(n_{eq}/a)$ of the dense stars in the galaxy, whose number is high. For instance, according to P.Kroupa [6] the most abundant population in galaxies are the dwarf stars (red, white and brown), whose gravity factor equals some units. To this one has to add the contribution of the high gravity factors of neutron stars and obscure bodies, each multiplied by the respective population.

- *The gravitational swallowing effect*

The "swallowing" of orbiting bodies, which is unknown to the classical gravitation (and has been found only recently between neutron star and luminous companion in binary systems), may be explained within the theory of the *local* gravitational constant $G$.

Eq.(11) shows that the effective gravity of a dense star $G(\delta) = (n_{eq}/a)G$ depends on its average density. Hence the gravity $G(\delta)$ of a *contracting* star increases and the orbiting bodies really flows towards it.

- *The fictitious nature of the relativistic black holes*

The contraction of the neutron star generates a high redshift due to the brake of the strong gravitational force (see eq.10), which reduces the photon energy
$$d(h\nu)/dr = -f(n,r)$$
giving the following gravitational redshift as observed at large distances
$$(12) \qquad z_g = (\nu_o - \nu)/\nu = e^\alpha - 1$$
where $\alpha = (n_{eq}/a)GM/c^2R = 9.34 \times 10^{17} \delta^{2/3}(1+\chi)G/c^2$.

This redshift reduces exponentially the photon energy as expected for any electromagnetic radiation following the thermodynamic laws. As a consequence only a small fraction of the radiation emitted from a neutron star may reach the observer. For instance, the exponent increases up to $\alpha \approx 32$ when the average density of a neutron star reaches $10^{16}$. This means that the star becomes *gradually* obscure.

In General relativity the *black holes* originate *abruptly* when the size of a contracting star equals the Schwarzschild radius, which marks a discontinuity of the relativistic redshift.

Due to the "gravitational mass" paradigm, the *black holes* undergo the embarrassing unlimited gravitational collapse, which was discussed in the past by R.Oppenheimer and others. Recently A.Loinger [15], going back critically to the Schwarzschild's work, showed that a *black hole* is a "mathematical" object deprived of the physical characteristics that a physical system needs to satisfy the thermodynamic laws.

Loinger analysed also [16] the properties of the old *dark body* studied by Michell and Laplace within the frame of the newtonian corpuscular theory of light, according to which a light's corpuscule emitted within the Schwarzschild's boundary cannot escape from the gravitational field.

These properties early migrate in the relativistic theory of the observed gravitational phenomena, producing the unphysical character of the black holes.

Conversely, in the undulatory theory of light the emitted photons cross the Schwarzschild's boundary, paying however the penalty of the strong reduction of energy.

This fundamental difference in conceiving the *obscure* bodies might appear manifest in the observation of the neutron star in binary systems.

**References**

1. L.Miller et al., Communication to Amer. Astron. Soc. Conference, January 2005 , S.Diego, Cal., *Nature News,* DOI 10.1038/050110-6

2. J.Maillard, T.Paumard, S.Stolovy, F.Rigaut ,"The nature of the galactic center source IRS 13 revealed by high spatial resolution in the infrared", *Astron. Astrophys.*, **423,**155-167 (2004)

3. F.Baganoff, M.Bautz, W.Brandt, G.Chartas et al., "Rapid X-ray flare from the direction of the supermassive black hole at the Galactic Centre", *Nature,* **413,**45-48 (2001)

4. R.Schoedel, T.Ott, R. Genzel, R. Hofmann et al., "A star in a 15.2-year orbit around the supermassive black hole at the centre of the Milky Way", *Nature,* **419,**694-696 (2002)

5. D.Figer, "An upper limit to the masses of stars", *Nature,* Vol.**434,**192-194 (2005)

6. P.Kroupa, "Stellar mass limited", *Nature News & Views,* Vol.**434,**148-149 (2005)

7. H.Van den Heuvel, G. Habets, "Observational low mass limit for black hole formation derived from massive X-ray binaries", *Nature,* **309,**598-600 (1984)

8. J.Greiner, J.Cuby, M.McCaughrean, "An unusually massive stellar black hole in the Galaxy", *Nature,* **414,**522-525 (2001)

9. R.Blandford, N.Gehrels, "Revisiting the black holes", *Physics Today,* June 1999

10. R.Larson, V.Bromm "Le prime stelle dell'universo", *Scientific American/Le Scienze*, Gennaio 2002

11. T.Heckman, G.Kauffmann, J.Brinchmann, S.Charlot et al., "Present-day growth of black holes and bulges: the Sloan Digital Sky Survey Perspective", *Astron. Journ.* **613,**109-118 (2004)

12. L.Krauss, M.Turner, "Rompicapo cosmico", *Scientific American/Le Scienze,* Novem.2004

13. M.Michelini, "The cosmic quanta paradigm fulfils the relativistic mechanics, improves the gravitation theory and originates the nuclear forces", *arXiv: physics/0509017* –Sept. 2005

14. A.Aguirre, C. Burgess. A. Friedland, D.Nolte, Astrophysical constrains on modifying gravity at large distances, *Class. Quantum Gravity*, **18,** 223-232 (2001)

15. A.Loinger, "On continued gravitational collapse", *arXiv: astro-ph/0001453* – Jan. 2000

16. A.Loinger, "On Michell-Laplace dark body", *arXiv: physics/0310058 v2* – Oct. 2003